\documentclass[conference]{IEEEtran}
\usepackage[dvips]{graphicx}

\usepackage{amssymb,amsmath}
\newtheorem{theorem}{Theorem}[section]
\newtheorem{lemma}[theorem]{Lemma}
\newtheorem{proposition}[theorem]{Proposition}
\newtheorem{corollary}[theorem]{Corollary}
\newcommand{\imp}[1]{\mathsf{imp}({#1}) }
\begin{document}

\title{Network Coding for Speedup in Switches}
\author{\authorblockN{MinJi Kim,
Jay Kumar Sundararajan, and Muriel M\'{e}dard}
\authorblockA{Laboratory for Information and Decision Systems\\
Massachusetts Institute of Technology\\
Cambridge, MA 02139, USA\\
Email: \{minjikim, jaykumar, medard\}@mit.edu}}

\maketitle

\begin{abstract}
We present a graph theoretic upper bound on speedup needed to
achieve 100\% throughput in a multicast switch using network coding.
By bounding speedup, we show the equivalence between network coding
and speedup in multicast switches - {\it i.e.} network coding, which
is usually implemented using software, can in many cases substitute
speedup, which is often achieved by adding extra switch fabrics.
This bound is based on an approach to network coding problems called
the ``enhanced conflict graph''. We show that the ``imperfection
ratio'' of the enhanced conflict graph gives an upper bound on
speedup. In particular, we apply this result to $K\times N$ switches
with traffic patterns consisting of unicasts and broadcasts only to
obtain an upper bound of $\min (\frac{2K-1}{K}, \frac{2N}{N+1})$.
\end{abstract}
\IEEEpeerreviewmaketitle

\section{Introduction}

The input-queued crossbar switch has been studied well, especially
in the context of unicast traffic. It is known that 100\% throughput
can be achieved \cite{uni100}, in the sense that as long as no input
or output is oversubscribed, traffic can be supported without
causing the queues to grow unboundedly. Therefore, to serve any
admissible unicast traffic, the input-queued crossbar switch does
not need to process packets faster than the input line
rate\footnote{The line rate of a switch is the rate at which packets
arrive or leave the switch at any one port.}, {\it i.e.} the switch
does not need \emph{speedup}.

The extension of the problem to multicast flows, however, is
intrinsically more difficult. Marsan \emph{et al.} \cite{marsan}
gave a characterization of the rate region achievable in a multicast
switch with fanout splitting\footnote{Fanout splitting is the
ability to serve partially a multicast cell to only a subset of its
destined outputs, and complete the service in subsequent time
slots.}, and also defined the optimal scheduling policy.
Interestingly, this work proved that unlike in the unicast case,
100\% throughput cannot be achieved for multicast flows in an
input-queued switch. In fact, the minimum speedup needed to achieve
100\% throughput grows unboundedly with the switch size.

In this paper, we discuss the same problem as \cite{marsan}, with
the following modification. The inputs are allowed to send linear
combinations of cells waiting in the queues, {\it i.e.}, they are
allowed to perform linear \emph{network coding} \cite{ahlswede} with
fanout splitting. The main contributions of this paper are:
\begin{enumerate}
\item We show that network coding can in many cases substitute speedup.
\item We provide a simple graph-theoretic upper bound on
speedup.
\item We prove an upper bound on speedup of $\min
\left(\frac{2K-1}{K}, \frac{2N}{N+1}\right)$ for an arbitrary
$K\times N$ switch with traffic pattern restricted to unicasts and
broadcasts only.
\end{enumerate}

Our work builds on the work by Sundararajan {\it et al.} \cite{cg},
\cite{codingSwitch}, which gave a graph-theoretic formulation of the
rate region of a multicast switch with intra-flow coding using
\emph{enhanced conflict graphs}. Given a traffic pattern, the
enhanced conflict graph $G = (V,E)$ is an undirected graph that
contains one vertex for every \emph{subflow}.\footnote{A flow is a
stream of packets that have common source and destination set. It is
represented by a 2-tuple $(i, J)$ consisting of the input $i$ and a
subset $J$ of outputs corresponding to the destination set of the
multicast stream. A subflow of flow $(i,J)$ is a part of a flow from
input $i$ that goes to a particular output in $J$. Therefore, a
subflow is a 3-tuple $(i, J, j)$ consisting of an input $i$, a
subset of outputs $J$ and one output $j \in J$.} An edge exists
between two vertices if they represent two subflows from the same
input or to the same output. Reference \cite{codingSwitch} shows
that the stable set polytope and the fractional stable set polytope
of an enhanced conflict graph are the rate region and the admissible
region of a network coding switch, respectively. This
graph-theoretic formulation helps us transform any given traffic
pattern into a conflict graph, and the properties of this graph can
be used to derive insight on the speedup required to achieve 100\%
throughput with coding. A similar graph-theoretic formulation was
used by Caramanis {\it et al.} in \cite{caramanis} in the context of
unicast traffic in Banyan networks.

Note that, for the case of fanout splitting without coding,
\cite{marsan} gave a characterization of the rate region as the
convex hull of certain modified departure vectors. However, a
graph-theoretic formulation of the same is not known. As a result,
it is significantly harder to characterize the speedup required to
achieve 100\% throughput for fanout splitting without coding.

The rest of the paper is organized as follows. Section \ref{def}
states preliminary definitions that will be used throughout this
paper. Section \ref{codingVspeedup} shows that network coding is
equivalent to speedup in a multicast switch to some extent. Section
\ref{speed_ratio} gives the relationship between speedup and
imperfection ratio of a conflict graph, which leads to our main
result - an upper bound on the minimum speedup required to achieve
100\% throughput in a multicast switch with coding. In Section
\ref{KxN}, we apply the result from Section \ref{speed_ratio} to a
$K\times N$ switch with traffic consisting only of unicasts and
broadcasts and give an upper bound on speedup of $\min
(\frac{2K-1}{K}, \frac{2N}{N+1})$. Finally, in Section
\ref{conclusion}, we summarize the contributions of this paper, and
present a conjecture on the actual minimum speedup needed to achieve
100\% throughput in a $2 \times N$ multicast switch with unicasts
and broadcasts only.

\section{Notation and Definitions}\label{def}

Let $G = (V, E)$ be an undirected graph with vertex set $V$ and edge
set $E$. A graph $G_1 = (V_1, E_1)$ is a subgraph of $G$ if $V_1
\subseteq V$ and $E_1 \subseteq E$. A graph $G_2 = (V_2, E_2)$ is an
\emph{induced subgraph} of $G$ if $V_2 \subseteq V$ and $(v_1 , v_2)
\in E_2$ if and only if $(v_1, v_2) \in E$. In addition, $G_2$ is
often denoted as $G(V_2)$ and is said to be induced by $V_2$. The
\emph{complement} of graph $G$ is a graph $\overline{G}$ on the same
vertex set $V$ such that two vertices of $\overline{G}$ are adjacent
if and only if they are not adjacent in $G$. The \emph{chromatic
number} of a graph $G$ is the smallest number of colors $\chi(G)$
needed to color the vertices of $G$ so that no two adjacent vertices
share the same color.

$G$ is a \emph{complete graph} if for every pair of vertices in $V$
there exists an edge connecting the two, and $V$ is called a
\emph{clique}. If for every pair of vertices in $V$ there is no edge
connecting the two, then $V$ is said to be a \emph{stable set}. $G$
is a \emph{hole} if it is a chordless cycle; $G$ is called an
\emph{odd hole} if it is a hole of odd length at least 5. $G$ is an
\emph{anti-hole} if its complement is a hole; $G$ is an \emph{odd
anti-hole} if its complement is an odd hole. $G$ is said to be
\emph{perfect} if for every induced subgraph of $G$, the size of the
largest clique equals the chromatic number.

\subsection{Stable Set Polytope}\label{ssp}

The \emph{stable set polytope} $STAB(G)$ of a graph $G$ is the convex hull
of the incidence vectors of the stable sets of the graph $G$. In this
section, we discuss how the stable set polytope of a conflict graph
can translate to the rate region of a switch.

Let $\mathbf{r} \in \mathbb{R}^f$ be the \emph{rate vector} of a traffic
pattern that has $f$ flows. Suppose that the total number of subflows
in the pattern is $m$. Then, the \emph{enhanced rate vector}
$e(\mathbf{r}) \in \mathbb{R}^m$ corresponding to $\mathbf{r}$ is
defined as:\vspace*{-0.2cm}
\[
\vspace*{-0.2cm}
e_{(i, J, j)}(\mathbf{r}) = \mathbf{r}_{(i, J)}, \text{ for all } j\in J.
\]
We use the enhanced rate vector as \emph{weights} for vertices of
the enhanced conflict graph.

A traffic pattern $\mathbf{r}$ is said to be \emph{achievable} if
there exists a switch schedule that can serve it; it is called
\emph{admissible} if no input or output is oversubscribed. We also
call the collection of all achievable and admissible vectors as the
\emph{achievable rate region} $\mathbf{R} \subseteq \mathbb{R}^f$
and \emph{admissible rate region} $\mathbf{A} \subseteq
\mathbb{R}^f$ respectively. For $\mathbf{r} \in \mathbf{R}$, we can
construct a switch schedule, which can be viewed as a time sharing
between valid switch configurations. In a conflict graph, a valid
switch configuration corresponds to a stable set, and a switch
schedule corresponds to a convex combination of stable sets of the
conflict graph $G$. Therefore, if a rate vector $\mathbf{r} \in
\mathbf{R}$, then $e(\mathbf{r})\in STAB(G) \subseteq \mathbb{R}^m$.

For a general graph $G$, a complete characterization of $STAB(G)$ in
terms of linear inequalities is unknown. However, several families
of necessary conditions are known. One example is the clique
inequalities\footnote{Clique inequalities of a graph say that the
total weight on the vertices of maximal cliques must not exceed 1.
In an enhanced conflict graph, the clique inequalities imply that no
input nor any output may be overloaded.}. The polytope described by
these conditions along with non-negativity
constraints\footnote{Non-negativity constraints of a graph say that
the weight on each vertex is non-negative.} is the \emph{fractional
stable set polytope} $QSTAB(G)$. In terms of the switch,
\cite{codingSwitch} shows that the clique inequalities of the
enhanced conflict graph correspond to the \emph{admissibility
conditions}. Therefore, if a rate vector $\mathbf{r} \in
\mathbf{A}$, then $e(\mathbf{r}) \in QSTAB(G) \subseteq
\mathbb{R}^m$.

Note that, for most graphs, $STAB(G) \subsetneq QSTAB(G)$, since the
clique inequalities are necessary but not sufficient conditions for
stable set polytope. Thus, the admissible
region is often a strict superset of the achievable rate region, which
implies that it is not possible to achieve 100\% throughput even
with fanout splitting and coding - we need speedup.

\subsection{Perfect Graph}\label{perfectgraph}

In this section, we focus on the properties of perfect graphs. We
first start by stating three well-known facts that characterize
perfect graphs.

\begin{theorem}\label{weak}\emph{(Weak Perfect Graph Theorem \cite{lovasz})
A graph $G$ is perfect if and only if its complement is perfect.}
\end{theorem}

\begin{theorem}\label{strong}\emph{(Strong Perfect Graph Theorem \cite{combinatorics}) A
graph $G$ is perfect if and only if it contains no odd hole and no
odd anti-hole.}
\end{theorem}

\begin{lemma}\label{replication}\emph{(Replication Lemma
\cite{lovasz}) Let $G = (V,E)$ be a perfect graph and $v\in V$.
Create a new vertex $v'$ and join it to $v$ and to all the neighbors
of $v$. Then, the resulting graph $G'$ is perfect.}
\end{lemma}

From Section \ref{ssp}, we have that $STAB(G) \subseteq QSTAB(G)$
for any graph with equality for perfect graphs only. This implies
that the admissible region $\mathbf{A}$ and the achievable rate
region $\mathbf{R}$ are the same if the enhanced conflict graph is
perfect. Thus, as given in {\it Corollary 1} from
\cite{codingSwitch}, if an enhanced conflict graph is perfect, then
speedup is not required to achieve 100\% throughput.

From this, we can observe that there is an intrinsic connection
between speedup and the ``perfectness'' of the enhanced conflict
graph. As a result, to compute the minimum speedup, it is helpful to
measure how perfect an enhanced conflict graph is. In this paper, we
use the \emph{imperfection ratio} introduced by Gerke and McDiarmid
\cite{gerke} as such a measure.

\subsection{Imperfection ratio}\label{impRatio}

In \cite{gerke}, the imperfection ratio $\imp{G}$ of graph $G$ is
defined as $\imp{G} = \min\{t : QSTAB(G) \subseteq t\ STAB(G)\}$.
As we noted in Section \ref{ssp}, in terms of a
switch, the admissible region $\mathbf{A}$ and the achievable region
$\mathbf{R}$ are projections of $QSTAB(G)$ and $STAB(G)$
respectively. Therefore, given the imperfection ratio $\imp{G}$ of an
enhanced conflict graph $G$, we have $\mathbf{A} \subseteq \imp{G}
\mathbf{R}$.

A useful bound on the imperfection ratio is presented in
\cite{gerke}, which we reproduce below.

\begin{proposition}\label{a/b} \emph{(Gerke and McDiarmid \cite{gerke})
For a graph $G$, if each vertex in $G$ can be covered $p$ times by a
family of $q$ induced perfect subgraphs, then $\imp{G} \leq
\frac{q}{p}$.}
\end{proposition}

\subsection{Speedup}\label{speedup}

A switch is said to have a \emph{speedup} $s$ if the switching
fabric can transfer packets at a rate $s$ times the incoming and
outgoing line rate of the switch. If we define a time slot to be the
reciprocal of the line rate, then this means the switching fabric
can go through $s$ configurations within one time slot. With this
definition, it is easy to see that a rate vector $\mathbf{r}$ is
achievable with speedup $s$ if and only if it is admissible and
$\frac{1}{s}\mathbf{r}$ is within the rate region.

Note that the admissible and achievable rates correspond to
$\mathbf{A}$ and $\mathbf{R}$ respectively. Then, $s_{\min} = \min\{
s\ |\ \mathbf{A} \subseteq s\ \mathbf{R}\}$
is the \emph{minimum speedup} required for the switch to achieve
all admissible rates, {\it i.e.} it is the minimum of all $s$ such that
$\frac{1}{s}\mathbf{r}$ is within the rate region for
all admissible rate vectors $\mathbf{r}$.

\section{Network Coding for Speedup}\label{codingVspeedup}

In this section, we show the equivalence between network coding
and speedup in multicast switches - {\it i.e.} network coding, which
is usually implemented using software, can in many cases substitute speedup,
which is often achieved by adding extra switch fabrics.

\begin{figure}[h!]
\begin{center}
\includegraphics[height=2.3cm]{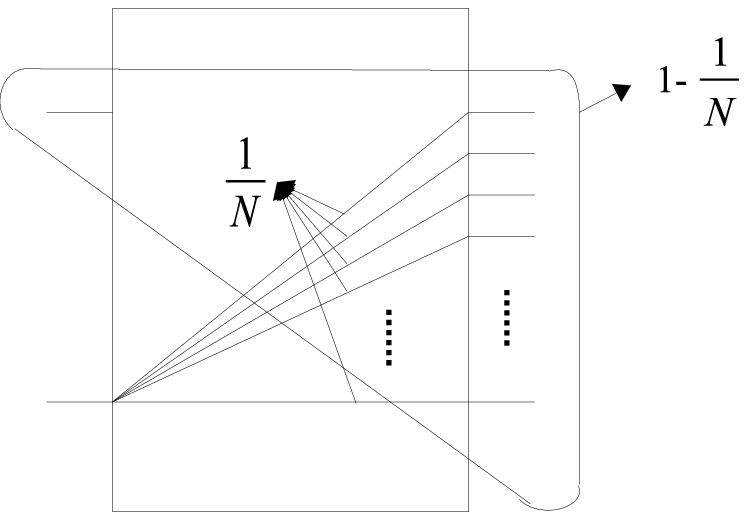}
\end{center}
\vspace{-.2cm} \caption{A traffic pattern which demonstrates the benefit of
coding}\label{benefit}
\end{figure}

In Figure \ref{benefit}, we show a special traffic pattern in a
$2\times N$ switch, which demonstrates the benefit of intra-flow
coding. At input 1, there is one broadcast flow with rate
$1-\frac{1}{N}$; at input 2, there is one unicast to each output
with rate $\frac{1}{N}$. Reference \cite{codingSwitch} shows that
this traffic is achievable if network coding with fanout splitting
is allowed; however, a speedup of $1.5-\frac{1}{N}$ is needed if
only fanout splitting is allowed. This example shows that network
coding is equivalent to a speedup of at least $1.5-\frac{1}{N}$.

\begin{figure}[h!]
\centerline{ \mbox{\includegraphics[height=2.3cm]{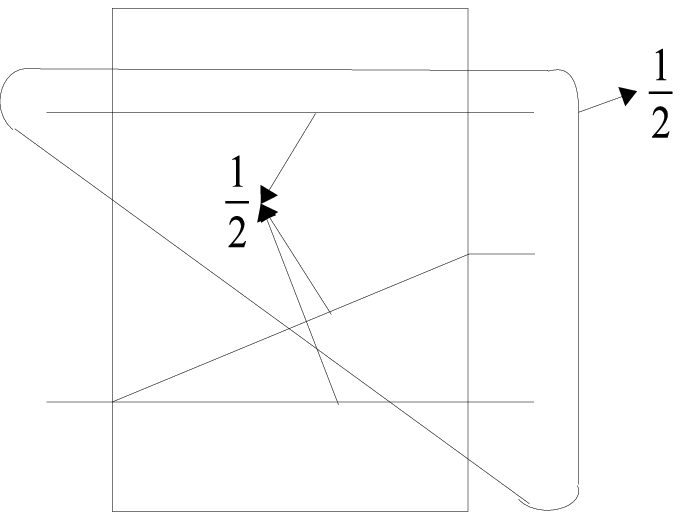}}
\mbox{\includegraphics[height=2.3cm]{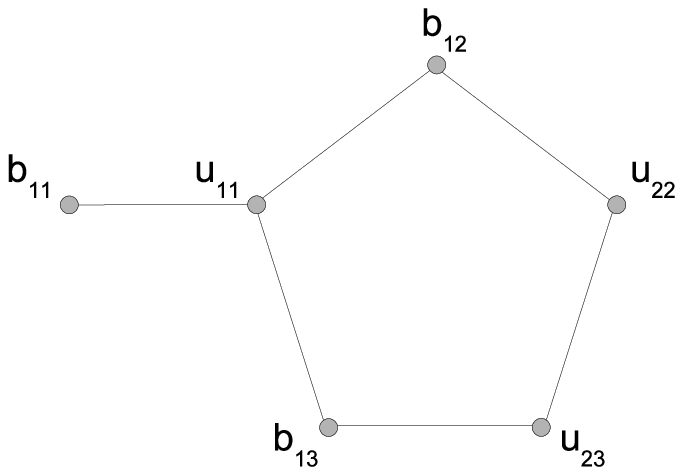}} } \caption{A traffic
pattern which requires speedup in a $2\times 3$ switch and its
enhanced conflict graph} \label{5/4} \vspace*{-.5cm}
\end{figure}

However, it is important to note that network coding cannot
completely replace speedup. As noted above in Figure \ref{benefit},
there are situations where network coding reduces speedup; however,
there are situations where speedup needed remains the same for with
and without network coding. For instance, in Figure \ref{5/4}, we
show a traffic pattern that requires speedup of 1.25 with or without
network coding. At input 1, there is a broadcast flow and a unicast
to output 1 with rate $\frac{1}{2}$ each; at input 2, there is one
unicast flow to each output 2 and 3 with rate $\frac{1}{2}$. In
Figure \ref{5/4}, we show that the enhanced conflict graph for this
traffic, where $u_{ij}$ represents the unicast flow from input $i$
to output $j$, and the vertex $b_{ij}$ represents the broadcast
subflow from input $i$ to output $j$. The enhanced conflict graph
contains an odd hole; therefore, it is not perfect.

Note that the traffic pattern in Figure \ref{5/4} gives a lower
bound on the speedup needed to achieve 100\% throughput in a
multicast switch using network coding. Therefore, $s_{\min} \geq
1.25$.

\section{Imperfection Ratio Bounds Speedup}\label{speed_ratio}
This section develops our main result, which relates speedup with
imperfection ratio. Note that, the definition of imperfection ratio
in Section \ref{impRatio} is very similar to that of minimum speedup
in Section \ref{speedup}. As a result, Corollary \ref{main} follows
from Proposition \ref{a/b}.

\begin{corollary}\label{main}
\emph{Given a traffic pattern, let $G$ be its enhanced conflict
graph and $s_{\min}$ be the minimum speedup required to achieve all
admissible rates. Then, $s_{\min} \leq \imp{G}$.}
\end{corollary}

Note that the converse of Corollary \ref{main} is not true. This is
because $\mathbf{A}$ and $\mathbf{R}$ are projections of $QSTAB(G)$
and $STAB(G)$ such that the subflows corresponding to the same
multicast flow have the same weight. As a result, $QSTAB(G)
\subseteq \imp{G} STAB(G)$ implies the $\mathbf{A} \subseteq \imp{G}
\mathbf{R}$, but $\mathbf{A} \subseteq s_{\min} \mathbf{R}$ may not
imply $QSTAB(G) \subseteq s_{\min} STAB(G)$.

\section{Bounds on Speedup for
$K\times N$ switch with unicasts and broadcasts}\label{KxN}

In this section, we apply Corollary \ref{main} to $K\times N$
switches using intra-flow coding with traffic patterns consisting of
unicasts and broadcasts only. We show that the minimum speedup
needed for 100\% throughput in this case is bounded by $\min
(\frac{2K-1}{K}, \frac{2N}{N+1})$. In this section, coding implies
intra-flow coding, since enhanced conflict graphs handle intra-flow,
not inter-flow, coding. The rest of this section is organized as
follows. First, we give a description of the enhanced conflict graph
for a $K\times N$ switch. In Section \ref{2k-1} and \ref{2n}, we
show the two bounds on speedup of $\frac{2K-1}{K}$ and
$\frac{2N}{N+1}$ respectively.

\subsection{Enhanced conflict graph for $K\times N$
switch}\label{ecgkn}

Consider traffic patterns which consist only of unicasts and a
broadcast per each input on a $K\times N$ switch. The basic idea
behind conflict graph is that vertices representing flows that
cannot be served simultaneously are adjacent. In such a case, the
enhanced conflict graph $G_{K,N} = (V,E)$ has the following
structure.

The vertex set $V = \left(\cup_{i \in [1, K]} U_i\right) \cup
\left(\cup_{i \in [1,K]} B_i\right)= \left(\cup_{j \in [1, N]}
U^o_j\right) \cup \left(\cup_{j \in [1,N]} B^o_j\right)$ where $U_i
= \{ u_{ij}\ |\ j\in [1,N]\}$\footnote{$j\in[1,N]$ means $j$ can be
integer from 1 to $N$.}, $B_i = \{ b_{ij}\ |\ j \in [1, N]\}$,
$U^o_j = \{ u_{ij}\ |\ i \in [1, K]\}$, and $B^o_j = \{ b_{ij} | i
\in [1, K]\}$. The vertex $u_{ij}$ represents the unicast flow from
input $i$ to output $j$, and the vertex $b_{ij}$ represents the
broadcast subflow from input $i$ to output $j$. Therefore, $U_i$ and
$U^o_j$ are collections of the unicast flows from input $i$ and to
output $j$ respectively. $B_i$ and $B^o_j$ are collections of the
subflows of the broadcast from input $i$ and to output $j$
respectively.

The edge set $E = \left( \cup_{i \in [1, K]} E^u_i\right) \cup
\left(\cup_{i \in [1, K]} E^b_i\right) \cup E^o$ where $E^u_i = \{
(u_{ij}, u_{ik})\ |\ j \ne k, j, k\in [1,N]\}$, $E^b_i = \{ (b_{ij},
u_{ik})\ |\ j, k\in [1, N]\}$, and $E^o = \cup_{i\in[1,N]} E^o_i$
where $E^o_i = \{(u_{ji}, u_{ki}), (b_{ji}, b_{ki}), (b_{ji},
u_{ki})\ |\ j \ne k, j,k \in[1,K]\}$. Each edge set represents a
different type of conflict. $E^u_i$ represents conflicts among
unicasts at input $i$; $E^b_i$ represents conflict between any
broadcast subflow and any unicast at input $i$; and $E^o_i$
represents conflicts among all flows and subflows at output $i$.

\begin{figure}[h]
\centering \vspace*{-.2cm}
\includegraphics[width = 8.5cm]{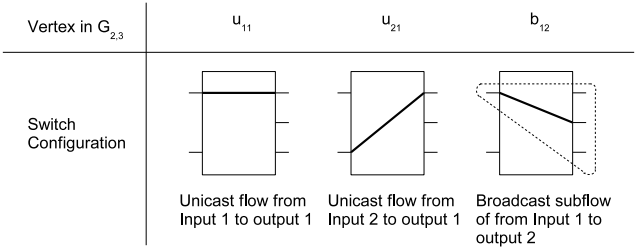}
\vspace{-.4cm} \caption{Switch configuration corresponding to
$u_{11}$, $u_{21}$, and $b_{12}$ in $G_{2,3}$}
\label{exConfiguration}\vspace*{-.1cm}
\end{figure}

It is important to note that each vertex in $G_{K,N}$ represents a
subflow in a $K\times N$ switch. For example, $u_{11}$ and $u_{21}$
corresponds to a unicast traffic to output 1 from input 1 and input
2 respectively. The vertex $b_{12}$ represents a partial service of
the broadcast from input 1 to output 2. In Figure
\ref{exConfiguration}, we show the switch configuration
corresponding to $u_{11}$, $u_{21}$, and $b_{12}$ in a $2\times 3$
switch.

\begin{figure}[h!]
\begin{center}
\includegraphics[width = 6cm]{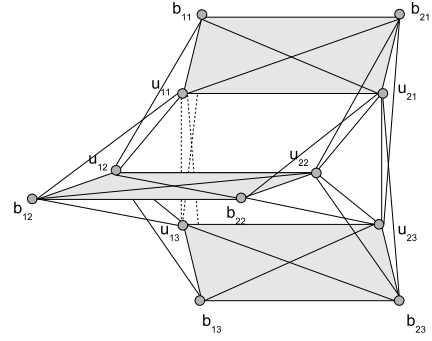}
\end{center}
\vspace{-.2cm} \caption{$G_{2,3}$ for a $2\times 3$ switch with
unicasts and broadcasts only} \label{ecg} \vspace{-.4cm}
\end{figure}

The intuition behind a conflict graph is that vertices which
represent flows that cannot be served simultaneously are adjacent.
As shown in \cite{codingSwitch}, if fanout splitting and network
coding are allowed, the switch can simultaneously serve two or more
subflows of the same broadcast flow and hence such subflows are not
adjacent to each other. For example, in Figure \ref{ecg}, there are
edges between $u_{11}$ and $b_{12}$, since they conflict at input 1,
and between $u_{11}$ and $u_{21}$, since they conflict at output 1;
however $u_{21}$ and $b_{12}$ are not adjacent, since they have
different input and output. Therefore, from the input perspective,
$G_{K,N}$ consists of $K$ induced complete subgraphs $G_{K,N}(U_i)$
for unicasts from each input $i$, and $K$ induced stable sets
$G_{K,N}(B_i)$ for broadcasts from each input $i$; from the output
perspective, $G_{K,N}$ consists of $2N$ induced complete subgraphs
$G_{K,N}(U^o_j)$ and $G_{K,N}(B^o_j)$ for unicasts and broadcast
subflows to each output $j$ respectively.

Here, we note that conflict graph of a $K\times N$ multicast switch
with unicasts and broadcasts traffic can be relaxed to that of
unicasts and single multicast per input. This relaxation just
removes vertices that represent broadcast subflows, which are not
part of the multicast flow, from the conflict graph. This cannot
hurt the ``perfectness'' of the conflict graph. Therefore, any upper
bound on the imperfection ratio of the conflict graph for unicasts
and broadcasts bounds that of unicasts and single multicast per
input.

\subsection{Speedup of $\frac{2K-1}{K}$}\label{2k-1}

In this Section, we give an upper bound on speedup for $K\times N$
switches. We present $2K-1$ induced perfect subgraphs of $G_{K,N}$
that cover $V$ $K$ times. Then, with Proposition \ref{a/b}, we
then have $\frac{2K-1}{K}$ as an upper bound for speedup.

\begin{lemma}\label{unicasts}
\emph{Let $G_u = G_{K,N}(\cup_{i \in [1,K]} U_i)$ be an induced subgraph
of $G_{K,N}$. Then $G_u$ is perfect.}
\end{lemma}
\begin{proof}
$G_u$ is an enhanced conflict graph for unicast traffic. One may
check that $G_u$ is a line graph of a bipartite graph, which is
known to be perfect \cite{combinatorics}.
\end{proof}

Lemma \ref{unicasts} also follows from the result in \cite{uni100}
which shows that 100\% throughput can be achieved in a input-queued
crossbar switch in the context of unicast traffic.

\begin{lemma}\label{UMM}
\emph{Let $G_i = G_{K, N}\left((\cup_{j \in [1,K]} B_j) \cup U_i\right)$
for some $i \in [1, K]$ be an induced subgraph of $G_{K,N}$. Then
$G_{i}$ is perfect.}
\end{lemma}
\begin{proof}
Assume that $G_i$ is not perfect. So it must have an odd hole or odd
anti-hole as an induced subgraph. Suppose it has an odd hole, say
$H$. In $G_i$, any broadcast subflow, except the ones from input
$i$, has no conflict on the input side. Suppose such a subflow were
part of $H$, then both its neighbors in $H$ will be due to output
side conflicts. But in that case, the two neighbors will themselves
conflict at the output, thereby forming a triangle. Since an odd
hole cannot contain a triangle, we conclude that $H$ cannot include
any $b_{jk}$, $j\neq i$.

This means $H$ must be an induced subgraph of $G_{K,N}(B_i \cup
U_i)$. However, $B_i$ induces a stable set, while $U_i$ induces a
clique. Therefore, $G_{K,N}(B_i \cup U_i)$ is a split
graph\footnote{A \emph{split graph} is a graph whose vertex set can
be partitioned into a stable set and a clique.} which is known to be
perfect \cite{combinatorics}. This contradiction shows that $G_i$
cannot contain an odd hole $H$.

Suppose $G_i$ contains an odd anti-hole, say $A$. This will happen
if and only if $\overline{G_i}$ contains an odd hole $H_A$. Note
that in $\overline{G_i}$, two vertices are connected if the
corresponding subflows do not conflict. Now, $H_A$ has to contain at
least one unicast, say $u_{ij}$, since the broadcasts by themselves
induce a perfect subgraph in $\overline{G_i}$ (they induce the
complement of a disjoint union of complete graphs, which is known to
be perfect \cite{combinatorics}). Now, $u_{ij}$ in $\overline{G_i}$
is adjacent to any $b_{i'j'}$, where $i\neq i'$ and $j\neq j'$. Let
$b_{pq}$ and $b_{p'q'}$ be vertices adjacent to $u_{ij}$ in $H_A$.
Then, using the definition of $\overline{G_i}$, we can infer that
$i\neq p\neq p'\neq i$ and $q=q'\neq j$. But this means, any vertex
that is adjacent to $b_{pq}$ is also adjacent to $b_{p'q'}$. Hence,
$H_A$ cannot be an odd hole.

This proves that $G_i$ is perfect.
\end{proof}

Using Lemmas \ref{unicasts} and \ref{UMM}, we derive our first
upper bound on speedup in $K\times N$ multicast switches with
traffic patterns consisting of unicasts and broadcasts only.

\begin{proposition}\label{2k-1/k}
\emph{$\imp{G_{K,N}} \leq \frac{2K-1}{K}$.}
\end{proposition}
\begin{proof}
Consider the following collection of induced subgraphs: $K-1$ copies
of $G_u$ from Lemma \ref{unicasts} and $G_i$ from Lemma \ref{UMM}
for all $i \in [1, K]$. We know that these subgraphs are all
perfect. In addition, these subgraphs cover each vertex in $v \in
G_{K,N}$ $K$ times. 
By Proposition \ref{a/b}, the claim follows.
\end{proof}

\subsection{Speedup of $\frac{2N}{N+1}$}\label{2n}

The proof idea in this section is similar to that of Section
\ref{2k-1}. We present $2N$ induced perfect subgraphs of $G_{K,N}$
that cover $V$ $N+1$ times, and then appeal to Proposition
\ref{a/b}. However, unlike Section \ref{2k-1}, here we change our
focus from the input to output.

\begin{lemma}\label{outunicasts}
\emph{Let $G^o_{1,i} = G_{K,N}(V_i)$ where $V_i = U^o_i \cup
\left(\cup_{j \in [1,N]} B^o_j\right)$ be an induced subgraph of
$G_{K,N}$. Then $G^o_{1,i}$ is perfect.}
\end{lemma}
\begin{proof}
Assume that $G^o_{1,i}$ is not perfect. So it must have an odd hole
or odd anti-hole as an induced subgraph. Suppose it has an odd hole,
say $H$. Since $U^o_i \cup B^o_i$ forms a complete graph (known to
be perfect), $H$ must contain vertices of $B^o_j$, $j\ne i$. Suppose
$b_{kj} \in B^o_j$ is part of $H$, then $H$ contains at least two
vertices of $B^o_j$. This is because, in $G^o_{1,i}$, $b_{kj}$ has
only one conflict on the input side; thus, neighbors of $b_{kj}$ are
$u_{ki}$ (input conflict) and $B^o_j$ (output conflict). However,
note that $B^o_j$ itself forms a complete graph, therefore $H$
contains at most two vertices of $B^o_j$. Thus, $b_{kj}$ and
$b_{k'j}$, $k \ne k'$ are in $H$. Then, $u_{ki}$ and $u_{k'i}$ are
in $H$. However, these four vertices form a cycle, thus $G^o_{1,i}$
cannot contain an odd hole $H$.

By the same argument as in the proof for Lemma \ref{UMM}, we can
show that $G^o_{1,i}$ cannot contain an odd anti-hole.
\end{proof}

\begin{lemma}\label{outUUM}
\emph{Let $G^o_{2,i} = G_{K,N}(V_i)$ where $V_i = B^o_i \cup
\left(\cup_{j \in [1,N]} U^o_j \right)$ be an induced subgraph of
$G_{K,N}$. Then, $G^o_{2,i}$ is perfect.}
\end{lemma}
\begin{proof}
$G^o_{2,i}$ is an enhanced conflict graph for unicast traffic in
addition to all broadcast subflows to output $i$. Consider $b_{1i}
\in B^o_i$ and $u_{1i} \in \cup_{i \in [1,K]} U_i$. In a $K\times N$
switch, $b_{1i}$ and $u_{1i}$ represent subflows from input 1 to
output $i$, and thus conflict with the same set of subflows, {\it i.e.}
neighbors of $u_{1i}$ are neighbors of $b_{1i}$. In addition,
$b_{1i}$ and $u_{1i}$ are in conflict. Therefore, by Replication
Lemma (Lemma \ref{replication}), we know that $G^o_{2,i}$ is perfect
if $G_{K,N}(V_i \setminus \{b_{1i}\})$ is perfect. We can apply this
argument repeatedly for each $b_{ji} \in B^o_i$, and deduce that if
$G_{K,N}(\cup_{j \in [1,N]} U^o_j)$ perfect then $G^o_{2,i}$ is
perfect. Note that from Lemma \ref{unicasts}, we know that the
enhanced conflict graph $G_u = G_{K,N}(\cup_{i \in [1,K]} U_i) =
G_{K,N}(\cup_{j \in [1,N]} U^o_j)$ for unicast traffic is perfect.
Therefore, $G^o_{2,i}$ is perfect.
\end{proof}

Now, using Lemmas \ref{outunicasts} and \ref{outUUM}, we can derive
an upper bound for speedup in $K\times N$ multicast switches with
traffic patterns consisting of unicasts and broadcasts only.

\begin{proposition}\label{2n/n+1}
\emph{$\imp{G_{K,N}} \leq \frac{2N}{N+1}$.}
\end{proposition}
\begin{proof}
Consider the following collection of induced subgraphs: $G^o_{1,i}$ and
$G^o_{2,i}$ for all $i \in [1, N]$. By Lemmas \ref{outunicasts}
and \ref{outUUM}, we know that these subgraphs are all perfect. In
addition, these subgraphs cover each vertex in $v \in G_{K,N}$ $N+1$
times. By Proposition \ref{a/b}, the claim follows.
\end{proof}

\section{Conclusion}\label{conclusion}
In this paper, we introduce a simple graph theoretic bound on
speedup needed to achieve 100\% throughput in a multicast network
coding switch using the concept of conflict graphs. We show that the
imperfection ratio of the conflict graph gives an upper bound on
speedup. We apply this result to $K\times N$ switches with traffic
patterns consisting of unicasts and broadcasts only to obtain an
upper bound of $\min (\frac{2K-1}{K}, \frac{2N}{N+1})$. For a
$2\times N$ switch, this gives a bound of 3/2 on speedup; however,
we conjecture that the actual speedup required to achieve 100\%
throughput in a $2\times N$ switch with traffic patterns consisting
of unicasts and broadcasts only is 5/4. We have verified this
conjecture using a computer for $N = 3, 4$ and 5.

In summary, by allowing network coding in multicast switches, we get
not only an insightful characterization of the speedup needed for
100\% throughput, but also a gain in speedup. We have shown that
network coding, which is usually implemented using software, can
substitute speedup, which is often achieved by adding extra switch
fabrics.


\section*{Acknowledgment}
This material is based upon research partly supported by Stanford
University under the Complex Network Infrastructures for
Communication and Power, Sponsor Award No. PY-1362; University of
California under DAWN: Dynamic Adhoc Wireless Networking, Sponsor
Award No. S0176938; Air Force Aerospace Research - OSR under the
Robust Self-Authenticating Network Coding, Sponsor Award No.
FA9550-06-1-0155; and DARPA ITMANET.

\end{document}